\pdfoutput=1 
\RequirePackage{fix-cm}
\documentclass[smallextended]{svjour3}       
\smartqed  
\usepackage{graphicx}
\usepackage{multicol}
\usepackage{epstopdf}
\DeclareGraphicsExtensions{.pdf,.jpeg,.png,.eps}

\usepackage{cite}
\usepackage[cmex10]{amsmath}

\usepackage{amsfonts}

\usepackage{latexsym}
\usepackage{algorithmicx}
\usepackage{algpseudocode}
\usepackage{listings}
\lstnewenvironment{Bench}[1][]
{\lstset{ %
		showlines=true,
		xleftmargin=5ex,
		firstnumber=1,
		language=C++,                
		basicstyle=\scriptsize\ttfamily,       
		numbers=left,                   
		numberstyle=\tiny,      
		stepnumber=1,                   
		numbersep=3pt,                  
		showspaces=false,               
		showstringspaces=false,         
		showtabs=false,                 
		tabsize=2,                      
		captionpos=b,                   
		breaklines=true,                
		breakatwhitespace=false,        
		escapeinside={\%*}{*)},          
		mathescape=true,
		columns=flexible,
		morekeywords={},
		escapeinside={(*@}{@*)},
		#1
	}}{}

\usepackage{booktabs}
\usepackage{multirow}
\usepackage{multicol}

\usepackage[caption=false,font=normalsize,labelfont=sf,textfont=sf]{subfig}
\usepackage{url}

\usepackage{xspace}

\usepackage{tkz-graph}
\usetikzlibrary {positioning}
\definecolor {processblue}{cmyk}{0.96,0,0,0}

\usepackage{color}

\newcommand{\DF}{Dataflow\xspace}
\newcommand{\df}{dataflow\xspace}
\newcommand{\BD}{Big Data\xspace}

\newcommand{\spark}{Spark\xspace}
\newcommand{\flink}{Flink\xspace}
\newcommand{\storm}{Storm\xspace}
\newcommand{\gtf}{TensorFlow\xspace}

\newcommand{\OP}[1]{{\tt #1}}

\usepackage[colorlinks]{hyperref}

\begin{document}

\title{A Comparison of Big Data Frameworks on a\\
	Layered \DF Model
}

\author{Claudia Misale \and Maurizio Drocco \and Marco Aldinucci \and Guy Tremblay}

\institute{C. Misale \and M. Drocco \and M. Aldinucci \at Computer Science 
Department,	University of Torino. Torino, Italy\\
    \email{\{misale, drocco, aldinuc\}@di.unito.it}           
	\and
	 G. Tremblay \at
	 D\'ept. d'Informatique, Universit\'{e} du Qu\'{e}bec \`{a} Montr\'{e}al. 
	 Montr\'{e}al, QC,  Canada \\
	 \email{tremblay.guy@uqam.ca}
}

\date{\today}

\maketitle

\begin{abstract}

In the world of Big Data analytics, there is a series of tools aiming at 
simplifying programming applications to be executed on clusters. Although each 
tool claims to provide better programming, data and execution models---for which only 
informal (and often confusing) semantics is generally provided---all share a 
common underlying model, namely, the \DF model. The \DF model we propose
shows how  various tools share the same expressiveness at different levels of 
abstraction.
The contribution of this work is twofold:
first, we show that the proposed model is (at least) as general as existing 
batch and streaming frameworks (e.g., \spark, \flink, \storm),
thus making it easier to understand high-level data-processing applications 
written in 
such frameworks. Second, we provide a layered model that can represent tools and 
applications following the \DF paradigm and we show how the analyzed 
tools fit in each level.
\end{abstract}

\section{Outline}
With the increasing number of \BD analytics tools, we witness a
continuous fight among implementors/vendors in demonstrating how their tools 
are better than others in terms of performances or expressiveness.
In this hype, for a user approaching \BD analytics  (even an educated computer
scientist), it might be difficult to have a clear picture of the
programming model underneath these tools and the expressiveness they provide 
to solve some user defined problem. With this in mind, we wanted to
understand the features those tools provide to the user in terms of
API and how they were related to parallel computing
paradigms. 

To provide some order in the world of \BD processing, in this paper we categorize 
some models and tools to  extract common features in their programming 
models.
We identified the \emph{\DF model} \cite{Lee:IEEE:P95} as the common model that better
describes all levels of abstraction, 
from the user-level API to the execution model. This model represents
applications as a  
directed graph of actors. In its ``modern'' reissue (aka macro-data
flow \cite{dataflow:pdp:12}),  it naturally models independent (thus parallelizable)
kernels starting from a graph of true data dependencies, where a kernel's
execution is triggered by data availability.

The Dataflow model is expressive enough to describe batch, 
micro-batch and streaming models that are implemented in most tools for 
\BD processing. Being all realized under the same common idea, we show how 
various \BD analytics tools share almost the same base concepts, differing mostly in their implementation 
choices. We instantiate the \DF model into a stack of layers where each 
layer represents a dataflow graph/model with a different meaning, describing a 
program from what the programmer sees down to the underlying, lower-level, 
execution model layer.
Furthermore, we put our attention to a problem arising from the high abstraction 
provided by the model that reflects into the examined tools. Especially when 
considering stream processing and state management, non-determinism may arise 
when processing one or more streams in one node of the graph, which is a 
well-known problem in parallel and distributed computing. Finally,
the paper also focus on high-level parallelism exploitation paradigms and
the correlation with Big Data tools  at the level of programming and execution
models.  

In this paper, we examine the following tools from a \df perspective:
Spark~\cite{zaharia_resilient_2012}, Storm~\cite{Anis:CoRR:storm:15}, 
Flink~\cite{flink-web}, and TensorFlow~\cite{tensorflow2015-whitepaper}.
We focus only on those tools since they are among the most famous and used ones nowadays. 
As far as we know, no previous attempt has been made to compare different \BD processing tools, at multiple levels of abstraction, under a common formalism.

The paper proceeds as follows. Section~\ref{sec:layered-model} describes the \DF model and how it can be exploited at three different abstraction levels. 
Section~\ref{sec:api} focuses on user-level API of the tools.
The various  levels of our layered model are discussed in
Sections~\ref{sec:semDF},  \ref{sec:execDF} and~\ref{sec:dpn}.
Then, Sections~\ref{sec:limit} and \ref{sec:skel} discuss some limitations of the \df model in capturing all the tool features and frames the programming model of the tools in a historical perspective.
Finally, Section~\ref{sec:conc} concludes the paper, also describing some future work.

\section{The \DF Layered Model}
\label{sec:layered-model}
By analyzing some well-known tools---\spark, \storm, \flink, and \gtf---we identified a common structure underlying all of them, based on the \DF model.
In Section~\ref{sec:dataflow} we review the \DF model of computation, as presented by Lee and Parks~\cite{Lee:IEEE:P95}.
In Section ~\ref{sec:stack}, we outline an architecture that can describe all these models at different levels of abstraction (Fig.~\ref{fig:stackmodel}, p.~\pageref{fig:stackmodel}), from the (top) user-level API to the (bottom-level) actual network of processes.
In particular, we show how the \DF model is general enough to subsume many different levels only by changing the semantics of actors and channels.

\subsection{The \DF Model}
\label{sec:dataflow}

\emph{\DF Process Networks} are a special case of \emph{Kahn Process Networks}, a model of computation that describes a program as a set of concurrent processes communicating with each other via FIFO channels, where reads are blocking and writes are non-blocking~\cite{Kahn74}.
In a \DF process network,
a set of \emph{firing rules} is associated with each process, called \emph{actor}.
Processing then consists of ``repeated firings of actors'', where an actor represents a \emph{functional} unit of computation over \emph{tokens}.
For an actor, to be \emph{functional} means that firings have no side effects---thus actors are stateless---and the output tokens are functions of the input tokens. The model can also be extended to allow stateful actors.

A \DF network can be executed mainly by two classes of execution, namely \emph{process-based} and  \emph{scheduling-based}---other models are flavors of these two.
The process-based  model is straightforward: each actor is represented by a process that 
communicates via FIFO channels.
In the scheduling-based model---also known as dynamic scheduling---a scheduler tracks the availability of tokens in input to actors and schedules enabled actors for execution; the atomic unit being scheduled is referred as a \emph{task} and represents the computation performed by an actor over a single set of input tokens.

\paragraph{Actors}
A \DF actor consumes input tokens when it ``fires'' and then produces output tokens; thus it repeatedly fires on tokens arriving from one or more streams. The function mapping input to output tokens is called the \emph{kernel} of an actor.\footnote{The \DF Process Network model also seamlessly comprehends the Macro \DF parallel execution model, in which each process executes arbitrary code. Conversely, an actor's code in a classical \DF \emph{architecture} model is typically a single machine instruction.
In the following, we consider \DF and Macro \DF to be equivalent models.}
A \emph{firing rule} defines when an actor can fire.
Each rule defines what tokens have to be available for the actor to fire.
Multiple rules can be combined to program arbitrarily complex firing logics (e.g., the \emph{If} node).

\paragraph{Input channels}
The kernel function takes as input one or more tokens from one or more
input channels when a firing rule is activated.
The basic model can be extended to allow for testing input channels for emptiness, to express arbitrary stream consuming policies (e.g., gathering from any channel: cf.\ Section~\ref{sec:limit}).

\paragraph{Output channels}
The kernel function places one or more tokens into one or more output channels when a firing rule is activated. Each output token produced by a firing can be replicated and placed onto each output channel (i.e., broadcasting) or sent to specific channels, in order to model arbitrarily producing policies (e.g., switch, scatter).

\paragraph{Stateful actors}
\label{par:state}
Actors with state can be considered like objects (instead of functions) with methods used to modify the object's internal state. Stateful actors is an extension that allows side effects over \emph{local} (i.e., internal to each actor) states.  It was shown by Lee and Sparks~\cite{Lee:IEEE:P95} that stateful actors can be emulated in the stateless \DF model by adding an extra feedback channel carrying the value of the state to the next execution of the kernel function on the next element of the stream and by defining appropriate firing rules.

\subsection{The \DF Stack}
\label{sec:stack}
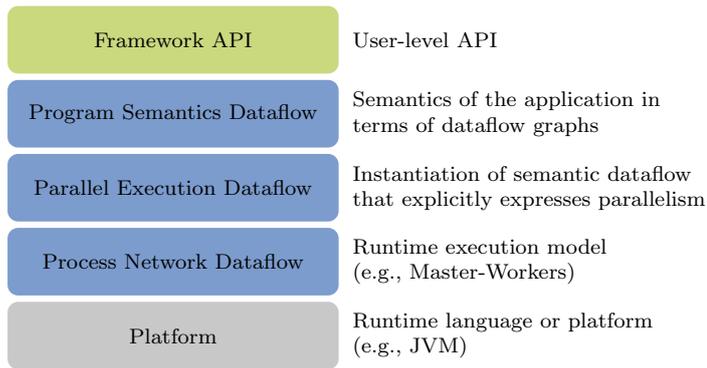
\begin{figure}
    \centering
\definecolor{mybluei}{RGB}{124,156,205}
\definecolor{myblueii}{RGB}{73,121,193}
\definecolor{mygreen}{RGB}{202,217,126}
\definecolor{mygray}{RGB}{200,200,200}

\pgfdeclarelayer{background}
\pgfsetlayers{background,main}

\tikzstyle{slayer}=[rounded corners,
text width=0.35\textwidth,
align=center,
minimum height=3em]

\tikzstyle{scapt}=[align=left,
text width=0.4\textwidth]

\begin{tikzpicture}[
node distance=2pt]

\node[slayer, fill=mygreen] (API) {Framework API};
\node[slayer, fill=mybluei, below=of API] (Sem) {Program Semantics Dataflow};
\node[slayer, fill=mybluei,below=of Sem] (Exec) {Parallel Execution Dataflow};
\node[slayer, fill=mybluei,below=of Exec] (Proc) {Process Network Dataflow};
\node[slayer, fill=mygray,below=of Proc] (Plat) {Platform};

\node[scapt,right=of API] () {User-level API};
\node[scapt, right=of Sem] () {Semantics of the application in terms of dataflow graphs};
\node[scapt, right=of Exec] () {Instantiation of semantic dataflow that explicitly expresses parallelism};
\node[scapt, right=of Proc] () {Runtime execution model\newline(e.g., Master-Workers)};
\node[scapt, right=of Plat] () {Runtime language or platform\newline(e.g., JVM)};

\end{tikzpicture}

    \caption{Layered model representing the levels of abstractions provided by the frameworks that were analyzed.}
    \label{fig:stackmodel}
\end{figure}
The layered model shown in Fig.~\ref{fig:stackmodel} presents five layers, where the three intermediate layers 
are \DF models with different semantics, as described in the paragraphs below.
Underneath these three layers is the \emph{Platform} level, that is, the runtime 
or programming 
language used to implement a given framework (e.g., Java and Scala in Spark), 
a level which is beyond the scope of our paper. 
On top is the \emph{Framework API} level, that describes the user API on top of the \DF 
graph, which will be detailed in Section~\ref{sec:api}.
The three \DF models in between are as follows.

\begin{itemize}
    \item \emph{Program Semantics \DF}:
    We claim the API exposed by any of the considered frameworks can be 
    translated into a \DF graph. The top level of our layered model captures this 
    translation: programs at this level represent the \emph{semantics}
    of data-processing applications in 
    terms of \DF graphs.
    Programs at this level do not explicitly express any form of parallelism: 
    they only express data dependencies (i.e., edges) among program 
    components (i.e., actors). This aspect is covered in Section 
    \ref{sec:semDF}.
    
    \item \emph{Parallel Execution \DF}:
    This level, covered in Section \ref{sec:execDF}, represents an 
    instantiation of the semantic \df{}s in 
    terms of processing elements (i.e., actors) connected by data channels 
    (i.e., edges). 
    Independent units---not connected by a channel---may execute in 
    parallel. 
    For example, a semantic actor can be replicated to express \emph{data 
        parallelism}, the execution model in which a given function is applied 
    to independent input data.
    
    \item \emph{Process Network \DF}:
    This level, covered in Section \ref{sec:dpn}, describes how the program is 
    effectively deployed and executed onto 
    the 
    underlying platform. Actors are concrete computing entities (e.g.,
    processes) 
    and edges are communication channels. The most common approach---used
    by 
    all the considered frameworks but \gtf---is for the actual network to be a 
    Master-Workers 
    task executor.
    In \gtf, processing elements are effectively mapped 
    to 
    threads and possibly distributed over multiple nodes of a cluster.
\end{itemize}

\section{The Frameworks' User APIs}
\label{sec:api}
Data-processing applications are generally divided into \emph{batch} vs.\
\emph{stream} processing. Batch programs process one or 
more \emph{finite} datasets to produce a resulting finite output dataset, whereas stream programs process possibly unbounded 
sequences of data, called \emph{streams}, doing so in an incremental manner. Operations over streams may also have to respect 
a total data ordering---for instance, to represent time ordering.

Orthogonally, we divide the frameworks' user APIs into two categories: \emph{declarative} and \emph{topological}. 
\spark, \flink, and \gtf belong to the first category---they
provide batch or stream processing in the form of operators over
collections or streams---whereas \storm belong to the
second one---it provides an API explicitly based on building graphs.

\subsection{Declarative Data Processing}
\label{sec:declarative}
This model provides as building blocks data collections and operations on those
collections. The data model follows domain-specific operators, for instance,
relational algebra operators that operate on data structured with the key-value 
model.

\emph{Declarative batch processing} 
applications are expressed as methods on objects representing
collections (\spark and \flink) or as functions on values (\emph{tensors}, in \gtf):
these are algebras on finite datasets, whose data can be ordered (as in tensors) or not (as in \spark/\flink multisets).
APIs with such operations are exposing a functional-like style.
Here are three examples of operations with their (multiset-based) semantics:%
\footnote{Here,  $\{\cdot\}$ denotes \emph{multisets} rather than sets.}
\begin{eqnarray}
\OP{groupByKey}(a) &=&  \{(k,\{v:(k,v)\in a\})\} \label{eq:gbk}\\
\OP{join}(a,b) &=&  \{(k,(v_a,v_b)):(k,v_a)\in a \land (k,v_b)\in b\} \label{eq:join}\\
\OP{map}\langle f \rangle(a) &=& \{f(v) : v \in a\} \label{eq:map}
\end{eqnarray}
The \OP{groupByKey} unary 
operation groups tuples sharing the same key (i.e., the first 
field of the tuple); thus it maps multisets of type $(K\times V)^*$ to 
multisets of type $(K\times V^*)^*$.
The binary \OP{join} operation merges two multisets by coupling values 
sharing the same key.
Finally, the unary higher-order \OP{map} operation applies the kernel 
function $f$ to each element in the input multiset.

\emph{Declarative stream processing}
programs are expressed in terms of an algebra on eventually unbounded data (i.e., stream as a whole) where data ordering eventually matters.
Data is usually organized in tuples having a key field used for example to express the position of each stream item with respect to a global order---a global timestamp---or to partition streams into substreams.
For instance, this allows expressing relational algebra operators and data grouping.
In a stream processing scenario, we also have to consider two important 
aspects: state and windowing; those are discussed in Section~\ref{sec:swi}.

\emph{Apache \spark}
implements batch programming with a set of operators, called 
\emph{transformations}, that are uniformly applied to whole datasets
called \emph{Resilient Distributed Datasets}
(RDD)~\cite{zaharia_resilient_2012}, which are immutable multisets.
For stream processing, Spark implements an extension through the Spark 
Streaming module, providing a high-level abstraction called 
\emph{discretized stream} or \emph{DStream}~\cite{Zaharia:2013:DSF:2517349.2522737}. Such
streams represent results in
continuous sequences of RDDs of the same type, called \emph{micro-batch}.
Operations over DStreams are ``forwarded'' to each RDD in the DStream, thus the semantics of operations over streams is defined in terms of batch processing according to the simple translation
$
\OP{op}({a}) = [\OP{op}(a_1), \OP{op}(a_2), 
\ldots]
$,
where $[\cdot]$ refers to a possibly unbounded ordered sequence, {$a = [a_1, a_2, \ldots]$ is a DStream, and each item 
$a_i$ is a micro-batch of type RDD.

\emph{Apache \flink}'s
main focus is on stream programming. The abstraction used is the 
DataStream, which is a representation of a stream as a single object. 
Operations are composed (i.e, pipelined) by calling operators on DataStream objects. 
\flink also provides the DataSet type for batch applications, that identifies a single immutable multiset---a stream of one element.
A \flink program, either for stream or batch processing, is a term from an algebra of operators over DataStreams or DataSets, respectively.
Stateful stream operators and iterative batch processing are discussed in Section~\ref{sec:swi}.

\emph{Google \gtf}
\label{par:gtf}
is a framework specifically designed for machine learning applications, where 
the data model consists of multidimensional arrays called \emph{tensors} and 
a program is a composition of operators processing
tensors.
A \gtf application is built as a functional-style expression, where
each sub-expression can be given an explicit name.
The \gtf programming model includes control flow operations and, notably, synchronization primitives (e.g., \emph{MutexAquire}/\emph{MutexRelease} for critical sections).
This latter observation implies \gtf exposes the underlying (parallel) execution model to the user which has to program the eventual coordination of operators concurring over some global state. 

\subsection{Topological Data Processing}
\label{sec:topological}
Topological programs are expressed as graphs, built by explicitly connecting processing nodes and specifying the code executed by nodes.

\emph{Apache \storm}
is a framework that only targets stream processing.
\storm's programming model is based on three key notions: \emph{Spouts}, \emph{Bolts}, and \emph{Topologies}.
A Spout is a source of a stream, that is (typically) connected to a data source or that can generate its own stream.
A Bolt is a processing element, so it processes any number of input streams and produces any number of new output streams.
Most of the logic of a computation goes into Bolts, such as functions, filters, streaming joins or streaming aggregations.
A Topology is the composition of Spouts and Bolts resulting in a network.
Storm uses \emph{tuples} as its data model, that is, named lists of values of arbitrary type.
Hence, Bolts are parametrized with per-tuple kernel code.
Each time a tuple is available from some input stream, the kernel code gets activated to work on that input tuple.
Bolts and Spouts are locally stateful, as we discuss in Section~\ref{sec:swi}, while no global consistent state is supported.
Yet, globally stateful computations can be implemented since the kernel code of Spouts and Bolts is arbitrary. However, eventual global state management would be the sole responsibility of the user, who has to be aware of the underlying execution model in order ensure program coordination among Spouts and Bolts.
It is also possible to define cyclic graphs by way of feedback channels connecting Bolts.

While \storm targets single-tuple granularity in its base interface, the Trident API is an abstraction that provides declarative stream processing on top of \storm. 
Namely, Trident processes streams as a series of micro-batches belonging to a stream considered as a single object.

\subsection{State, Windowing and Iterative Computations}
\label{sec:swi}

Frameworks providing \emph{stateful} stream processing make it possible to express modifications (i.e., side-effects) to the system state that will be visible at some future point.
If the state of the system is \emph{global}, then it can be accessed by all system components.
For example, \gtf mutable variables are a form of global state, since they can be attached to any processing node.
On the other hand, \emph{local} states can be accessed only by a single system component.
For example, the \OP{mapWithState} functional in the \spark Streaming API realizes a form of local state, in which successive executions of the functional see the modifications to the state made by previous ones.
Furthermore, state can be partitioned by shaping it as a tuple space, following, for instance, the aforementioned key-value paradigm.
With the exception of \gtf, all the considered frameworks provide local key-value states.

\emph{Windowing} is another concept provided by many stream processing frameworks.
A \emph{window} is informally defined as an ordered subset of items extracted from the stream.
The most common form of windowing is referred as \emph{sliding window} and it is characterized by the size (how many elements fall within the window) and the sliding policy (how items enter and exit from the window).
\spark provides the simplest abstraction for defining windows, since they are just micro-batches over the DStream abstraction where it is possible to define only the window length and the sliding policy.
\storm and \flink allow more arbitrary kinds of grouping, producing windows of Tuples and WindowedStreams, respectively.
Notice this does not break the declarative or topological nature of the considered frameworks, since it just changes the type of the processed data.
Notice also windowing can be expressed in terms of stateful processing, by considering window-typed state.

Finally, we consider another common concept in batch processing, namely \emph{iterative} processing. In \flink, iterations are expressed as the composition of arbitrary DataSet values by iterative operators, resulting in a so-called \emph{IterativeDataSet}.
Component DataSets represent for example \emph{step functions}---executed in each iteration---or termination condition---evaluated to decide if iteration has to be terminated.
\spark's iteration model is radically simpler, since no specific construct is provided to implement iterative processing. Instead, an RDD (endowed with transformations) can be embedded into a plain sequential loop.
Finally, \gtf allows expressing conditionals and loops by means of specific control flow operators such as \emph{For}, similarly to \flink.

\section{Program Semantics Dataflow}
\label{sec:semDF}
This level of our layered model provides a \DF representation of the program 
semantics. Such a model describes the application using operators and 
data dependencies among them, thus creating a topological view common to all frameworks.
This level does not explicitly express
parallelism: instead, parallelism is \emph{implicit} through the data dependencies among actors (i.e., among 
operators), so that operators which have no direct or indirect dependencies can be executed concurrently.

\subsection{Semantic \DF Graphs}
A semantic \DF graph is a pair $G\,=\,\langle V,\,E\,\rangle$ where actors $V$ represent operators, channels $E$ represent data dependencies among operators and tokens represent data to be processed.
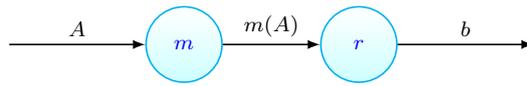
\begin{figure}
	\centering
	\begin {tikzpicture}[-latex, auto, node distance=2.3 cm, on grid ,
	semithick, state/.style ={circle, top color=white, bottom color=processblue!20,
		draw, processblue, text=blue, minimum width=1 cm}]
	\node[state] (M) {$m$};
	\node[state] (R) [right=of M] {$r$};
	\node (A) [left of=M, coordinate] {A};
	\node (B) [right of=R, coordinate] {b};
	\path (A) edge node {$A$} (M);
	\path (M) edge node {$m(A)$} (R);
	\path (R) edge node {$b$} (B);
\end{tikzpicture}
\caption{Functional Map and Reduce \df expressing data dependencies.} 
\label{fig:graph:mr-semantic}
\end{figure}
For instance, consider a map function~$m$ followed by a reduce function~$r$ on a collection $A$ and its result $b$, represented as the functional composition $ b=r(m(A))$.
This is represented by the graph in Fig.~\ref{fig:graph:mr-semantic}, which represents the semantic \df of a simple map-reduce program.
Notice the user program translation into the semantic \df can be subject to further optimization.
For instance, two or more non-intensive kernels can be mapped onto the same actor to reduce resource usage.

\begin{figure}
    \centering
    \subfloat[{\small \spark DAG}]{{\includegraphics[width=0.35\linewidth]{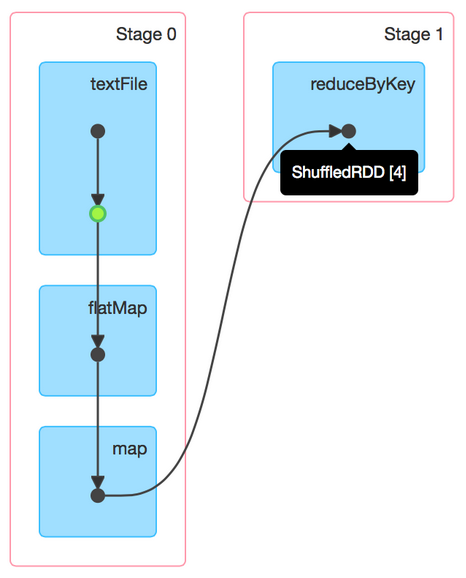}}}
    \qquad
    \subfloat[{\small \flink JobGraph}]{{\includegraphics[width=0.21\linewidth]{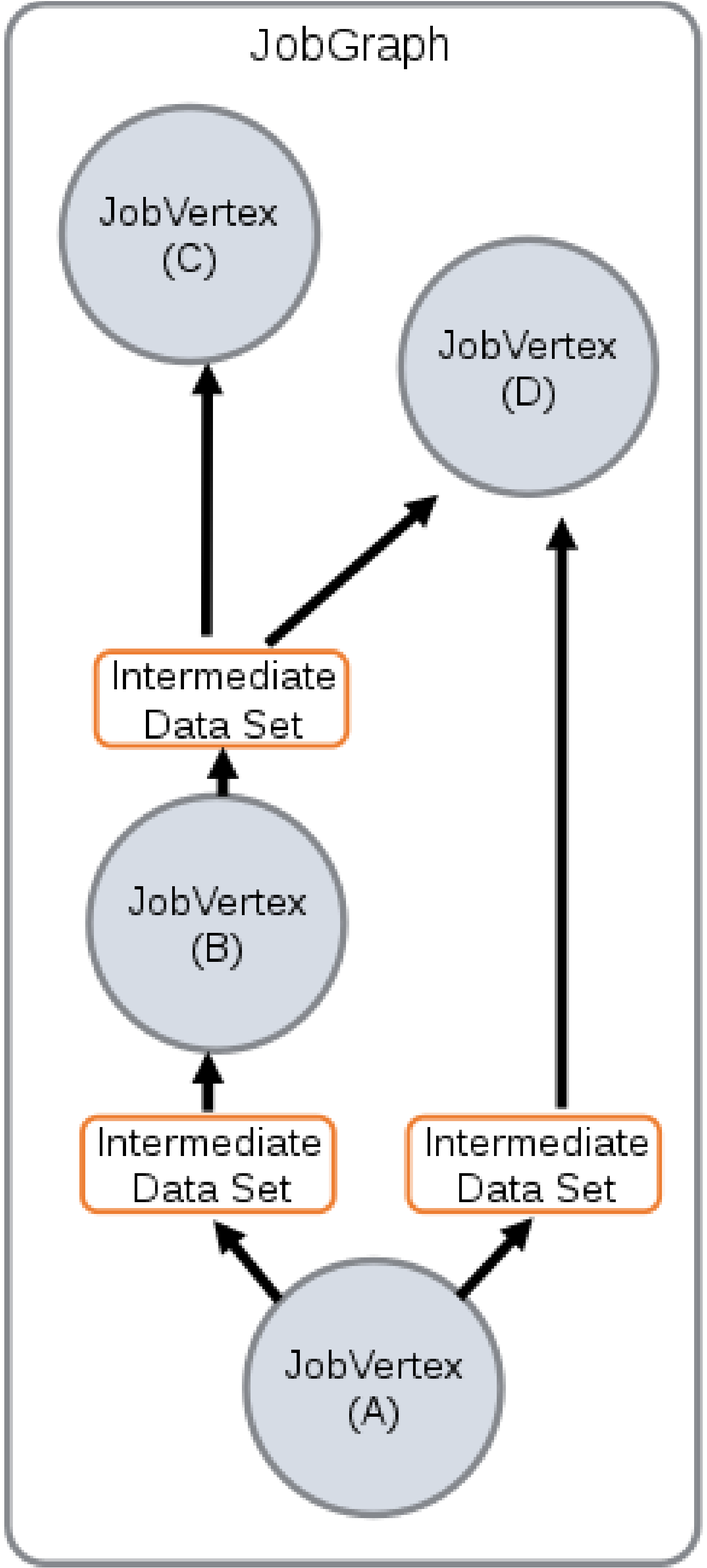}}}
    \qquad
    \subfloat[{\small \gtf graph}]{{\includegraphics[width=0.25\linewidth]{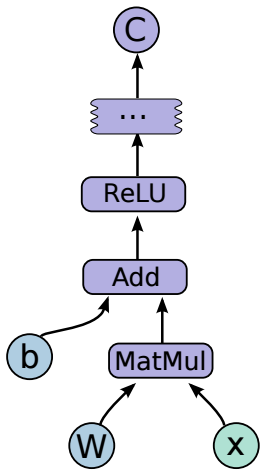}}}
    \caption{\spark DAG of the WordCount application (a). A \flink JobGraph (b). A \gtf application graph, adapted from~\cite{tensorflow2015-whitepaper} (c). }
    \label{fig:spark-flink-DAG}
\end{figure}
Notably, the \DF representation we propose is adopted by the considered frameworks as a pictorial representation of applications.
Fig. \ref{fig:spark-flink-DAG}(a) shows the semantic \df---called application DAG in \spark---related to the WordCount application, having as operations (in order): 1.~read from text file; 2.~a \OP{flatMap} operator splitting the file into words; 3.~a \OP{map} operator that maps each word into a key-value pair $(w,1)$; 4.~a \OP{reduceByKey} operator that counts occurrences of each word in the input file. 
Note that the DAG is grouped into \emph{stages} (namely, Stages~0 and~1), which divide \emph{map} and \emph{reduce} phases. This distinction is related to the underlying parallel execution model and will be covered in Section~\ref{sec:execDF}.
\flink also provides a semantic representation---called JobGraph or \emph{condensed view}--- of the application, consisting of operators (JobVertex) and intermediate results (IntermediateDataSet, representing data dependencies among operators).
Fig.~\ref{fig:spark-flink-DAG}(b) presents a small example of a JobGraph.
Finally, Fig.~\ref{fig:spark-flink-DAG}(c) is a \gtf example from~\cite{tensorflow2015-whitepaper}.
A node of the graph represents a tensor operation, which can be also a data generation one (nodes $W$, $b$, $x$).
Each node has firing rules that depend on the kind of incoming tokens.
For example,  \emph{control dependencies} edges can carry synchronization tokens: the target node of such edges cannot execute until all appropriate synchronization signals have been received. 

\subsection{Tokens and Actors Semantics}
\label{sec:granularity}

Although the frameworks provide a similar semantic expressiveness, some differences are visible regarding the meaning of tokens flowing across channels and how many times actors are activated. 

When mapping a \spark program, tokens represent RDDs and DStreams for batch and stream processing respectively.
Actors are operators---either transformations or actions in \spark nomenclature---that transform data or return  values (in-memory collection or files).
Actors are activated only once in both batch and stream processing, since each collection (either RDD or DStreams) is represented by a single token.
For \flink the approach is similar: actors are activated only once in all scenarios except in iterative algorithms, as we discuss in Section~\ref{sec:swiDF}.
Tokens represent DataSets and DataStreams that identify whole datasets and streams respectively.
For \gtf the same mapping holds: operators are mapped to actors that take in input single tokens representing Tensors (multi-dimensional arrays).
Actors are activated once except for iterative computations, as in \flink.
\storm is different since tokens represent a single item (called Tuple) of the stream. Consequently, actors, representing (macro) dataflow operators, are activated each time a new token is available.

From the discussion above, we can note that \storm's actors follow a \emph{from-any} policy for consuming input tokens, while the other frameworks follow a \emph{from-all} policy as in the basic \DF model. In all the considered frameworks, output tokens are broadcast onto all channels going out of a node. 

\subsection{Semantics of State, Windowing and Iterations}
\label{sec:swiDF}
In Section~\ref{sec:swi} we introduced stateful, windowing and iterative processing as convenient tools provided by the considered frameworks.

From a \DF perspective, stateful actors represent an extension to the basic model---as we sketched in Section~\ref{sec:dataflow}---only in case of global state.
In particular, globally-stateful processing breaks the functional nature of the basic \DF model, inhibiting for instance to reason in pure functional terms about program semantics (cf.\ Section~\ref{sec:limit}).
Conversely, locally-stateful processing can be emulated in terms of the pure \DF model, as discussed in~\cite{Lee:IEEE:P95}.
As a direct consequence, windowing is not a proper extension since windows can be stored within each actor's local state~\cite{dem:ijpp:2016}.
However, the considered frameworks treat windowing as a primitive concept.
This can be easily mapped to the \DF domain by just considering tokens of proper types.

Finally, iterations can be modeled by inserting loops in semantic \df{s}.
In this case, each actor involved in an iteration is activated each time a new token is available and the termination condition is not met.
This implementation of iterative computations is similar to the hierarchical actors of Lee~\&~Parks~\cite{Lee:IEEE:P95}, used to encapsulate subgraphs modeling iterative algorithms.

\section{Parallel Execution Dataflow}
\label{sec:execDF}
This level represents parallel implementations of semantic \df{s}.
As in the previous section, we start by introducing the approach and then we describe how the various frameworks instantiate it and what are the consequences this brings to the runtime.

The most straightforward source of parallelism comes directly from the \DF model, namely, independent actors can run in parallel.
Furthermore, some actors can be replicated to increase parallelism by making replicas work over a \emph{partition} of the input data---that is, by exploiting full \emph{data parallelism}.
This is the case, for instance, of the \OP{map} operator defined in Section~\ref{sec:declarative}.
Both the above schemas are referred as \emph{embarrassingly parallel} processing, since there are no dependencies among actors.
Note that introducing data parallelism requires partitioning input tokens into sub-tokens, distributing those to the various worker replicas, and then aggregating the resulting sub-tokens into an appropriate result token---much like \OP{scatter}/\OP{gather} operations in message passing programs.
Finally, in case of dependent actors that are activated multiple times, parallelism can still be exploited by letting tokens ``flow'' as soon as each activation is completed.
This well-known schema is referred as  \emph{stream}/\emph{pipeline} parallelism.

\begin{figure}
\centering
\begin{tikzpicture}[edge from 
parent/.style={draw,<-,>=latex},level/.style={sibling distance=41mm/#1},level 
distance=1cm, scale=0.8, transform shape]
\node [circle,draw,top color=white, bottom color=blue!50] (r0){$r$}
child {node [circle,draw,top color=white, bottom color=blue!30] (r1) {$r$}
	child {node [circle,draw,top color=white, bottom color=blue!30] (r2) {$r$}
		child {node [rectangle,draw,top color=white, bottom color=green!40] {$m$}}
		child {node[rectangle,draw,top color=white, bottom color=green!40]{$m$}}
	}
	child {node [circle,draw,top color=white, bottom color=blue!30] (r3) {$r$}
		child {node [rectangle,draw,top color=white, bottom color=green!40] {$m$}}
		child {node[rectangle,draw,top color=white, bottom color=green!40]{$m$}}
	}
}
child {node [circle,draw,top color=white, bottom color=blue!30] (r4) {$r$}
	child {node [circle,draw,top color=white, bottom color=blue!30] (r5) {$r$}
		child {node [rectangle,draw,top color=white, bottom color=green!40] {$m$}}
		child {node[rectangle,draw,top color=white, bottom color=green!40]{$m$}}
	}
	child {node [circle,draw,top color=white, bottom color=blue!30] (r6) {$r$}
		child {node [rectangle,draw,top color=white, bottom color=green!40] {$m$}}
		child {node[rectangle,draw,top color=white, bottom color=green!40]{$m$}}
	}
};
\end{tikzpicture}

\caption{MapReduce execution \df with maximum level of parallelism reached by eight $map$ instances.}
\label{fig:execMR}
\end{figure}
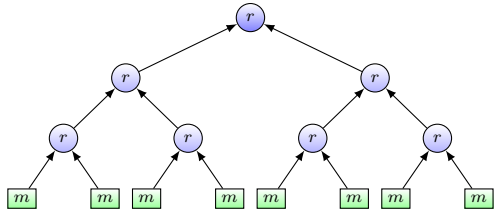
Fig.~\ref{fig:execMR} shows a parallel execution \df for the MapReduce semantic \df in Fig.~\ref{fig:graph:mr-semantic}.
In this example, the dataset $A$ is divided in 8 independent partitions and the map function $m$  is executed by 8 actor replicas;  the reduce phase is then executed in parallel by actors enabled by the incoming tokens (namely, the results) from their ``producer'' actors. 

\begin{figure}
    \centering
    \subfloat[Spark Execution DAG]{{\includegraphics[width=.3\linewidth]{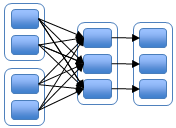} }}
    \qquad
    \subfloat[\flink Execution Graph]{{\includegraphics[width=.5\linewidth]{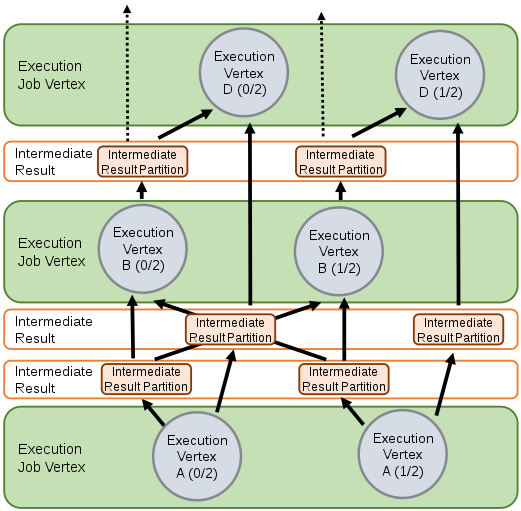}}}
    \caption{Parallel execution \df of a simple Map/Reduce application in \spark and \flink.}
    \label{fig:spark-flink}
\end{figure}
{\spark}
identifies its parallel execution \df by a DAG such as the one shown in 
Fig.~\ref{fig:spark-flink}(a), which is the input of the DAG Scheduler entity. 
This graph illustrates two main aspects: 
first, the fact that many parallel instances of actors are created for each function 
and, second, the actors are grouped into the so called 
\emph{Stages} that are executed in parallel if and only if there is no dependency among them. Stages can be considered as the hierarchical actors in~\cite{Lee:IEEE:P95}.
The actors grouping in stages brings another strong consequence, derived from the implementation of the Spark runtime: each stage that depends on one or more previous stages has to wait for their completion before execution.
The depicted behavior is analogous to the one encountered in the Bulk Synchronous Parallelism paradigm (BSP)~\cite{Valiant:BSP:1990}.
In a BSP algorithm, as well as in a Spark application, a computation proceeds 
in a series of global \emph{supersteps} consisting in: 1)~Concurrent 
computation, in which each actor executes its business code on its own 
partition of data; 2)~Communication, where actors exchange data between 
themselves if necessary (the so called \emph{shuffle} phase); 3)~Barrier 
synchronization, where actors
wait until all other actors have reached the same barrier.

{\flink}
transforms the JobGraph (Fig.~\ref{fig:spark-flink-DAG}(b)) into the ExecutionGraph~\cite{Carbone:Flink:FEHT:15} 
(Fig.~\ref{fig:spark-flink}(b)), in which the 
JobVertex (a hierarchical actor) is an abstract vertex containing a certain number of 
ExecutionVertexes (actors), one per parallel sub-task.
A key difference compared to the Spark execution graph is that a dependency does not represent a barrier among actors or hierarchical actors: instead, there is effective tokens pipelining and actors can be fired concurrently.
This is a natural implementation for stream processing, but in this case, since the runtime is the same, it applies to batch processing applications as well.
Conversely, iterative processing is implemented according to the BSP approach: one evaluation of the step function on all parallel instances forms a 
\emph{superstep} (again a hierarchical actor), which is also the granularity of synchronization; all parallel tasks of an iteration need to complete the superstep before the next one is initiated, thus behaving like a \emph{barrier} between iterations.

{\gtf}
replicates actors implementing certain operators (e.g., tensor multiplication) on tensors (input tokens).
Hence,
each actor is a data-parallel actor operating on intra-task 
independent input elements---here, multi-dimen\-sional arrays (tensors).
Moreover, iterative actors/hierarchical actors (in case of cycles on a subgraph) are implemented with tags 
similar to the MIT Tagged-Token \df machine~\cite{Arvind:1990:EPM}, where the iteration state is identified by a tag and independent iterations are executed in parallel.
It is worthwhile to remark \gtf differs from \flink in the execution of iterative actors: in \gtf an input can enter a loop iteration whenever it becomes available, while \flink imposes a barrier after each iteration.

{\storm}
creates an environment for the execution dataflow similar to the 
other frameworks. Each 
actor is replicated to increase the inter-actor parallelism and each group of replicas is 
identified by the name of the Bolt/Spout of the semantics dataflow they 
originally belong to, thus instantiating a hierarchical actor. Each of these actors (actors group) represents data parallel 
tasks without dependencies. Since Storm is a stream processing framework, 
pipeline parallelism is exploited. Hence, while an actor is processing a 
token (tuple), an upstream actor can process the next token concurrently, 
increasing both data parallelism within each actors group and task parallelism 
among groups.


\newcommand{\COLA}{p{0.13\linewidth}}
\newcommand{\COLB}{p{0.22\linewidth}}
\newcommand{\COLC}{p{0.28\linewidth}}

\begin{table}
{\footnotesize
\begin{tabular}{\COLA\COLB\COLB\COLC}
\toprule
 & \textbf{Spark} &\textbf{Flink} &\textbf{TensorFlow}\\
\midrule
{\bf Graph specification} 
     &
     Implicit, OO-style chaining of transformations
     &
     Idem
     &
     Implicit, Prefix operator with arguments\\
\midrule
\textbf{ DAG} 
     &
     Join operation 
     &
     Idem
     &
     N-ary operators and/or results\\
\midrule
\textbf{Tokens}
     &
     RDD
     &
     DataSet
     & 
     Tensor\\
\midrule
\textbf{Nodes} 
     &
     Transformations from RDD to RDD
     &
     Transformations from DataSet to DataSet
     &
     Transformations from Tensor to Tensor\\
\midrule
\textbf{Parallelism}
     &
     Data parallelism in transformations +\newline
     Inter-actor, task parallelism, limited by per-stage BSP
     &
     Data parallelism in transformations +\newline Inter-actor task parallelism
     &
     Idem + Loop parallelism\\
\midrule
\textbf{Iteration}
     &
     Using \emph{repetitive sequential executions} of the graph
     &
     Using {\tt iterate \& iterateDelta}
     &
     Using control flow constructs\\
\bottomrule
\end{tabular}
}
\caption{Batch processing.\label{tab:comparison batch}}
\end{table}

\begin{table}
{\footnotesize
\begin{tabular}{\COLA\COLB\COLB\COLC}
\toprule
 &\textbf{Spark} &\textbf{Flink} &\textbf{Storm}\\
\midrule
\textbf{Graph specification}&
     Implicit, OO-style chaining of transformations &
     Idem&
     Explicit, Connections between \emph{bolts}\\
\midrule  
\textbf{DAG}&
     Join operation &
     Idem &
     Multiple incoming/outgoing connections\\
\midrule 
\textbf{Tokens}&
     DStream  &
     DataStream & 
     Tuple (fine-grain)\\
\midrule
\textbf{Nodes} 
     &
     Transformations from DStream to DStream
     &
     Transformations from Data\-Stream to DataStream
     &
     Stateful with ``arbitrary'' emission of output tuples\\
\midrule
\textbf{Parallelism}
     &
     Analogous to \spark Batch parallelism
     &
     Analogous to \flink Batch parallelism+\newline
     Stream parallelism between stream items
     &
     Data parallelism between different bolt instances +\newline
     Stream parallelism between stream items by bolts\\
\bottomrule
\end{tabular}
}
\caption{Stream processing.\label{tab:comparison streaming}}
\end{table}

Summarizing, in sections \ref{sec:semDF} and \ref{sec:execDF} we showed how the considered frameworks can be compared under the very same model from both a semantic and a parallel implementation perspective.
The comparison is summarized in Tables~\ref{tab:comparison batch} and \ref{tab:comparison streaming} for batch and streaming processing, respectively.

\section{\DF Process Network}
\label{sec:dpn}

This layer shows how the program is effectively executed, following the process and scheduling-based categorization described in Section~\ref{sec:dataflow}.

\subsection{Scheduling-based Execution}
In \spark, \flink and \storm, the resulting process network dataflow follows the Master-Workers pattern, where actors from
previous layers are transformed into tasks. 
Fig.~\ref{fig:common-mw}(a) shows a representation of the Spark Master-Workers
runtime. We will use this structure also to examine Storm and 
Flink, since the pattern is similar for them: they differ only in how tasks are 
distributed among workers and how the inter/intra-communication between actors is managed. 

\begin{figure}
    \centering
    \subfloat[Master-Workers]
    {{\includegraphics[width=0.48\linewidth]{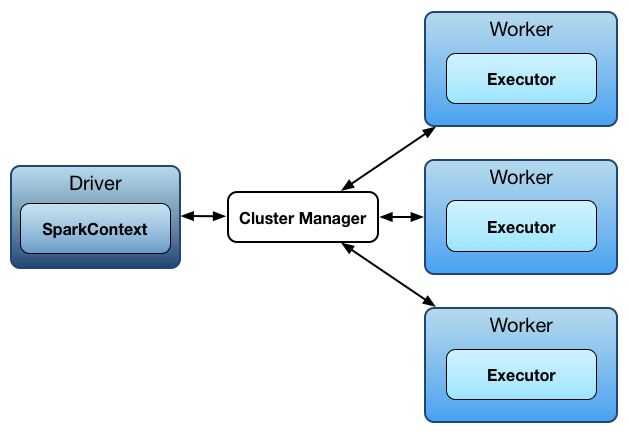}}}
    ~
    \subfloat[Worker hierarchy]{{	
            \includegraphics[width=0.42\linewidth]{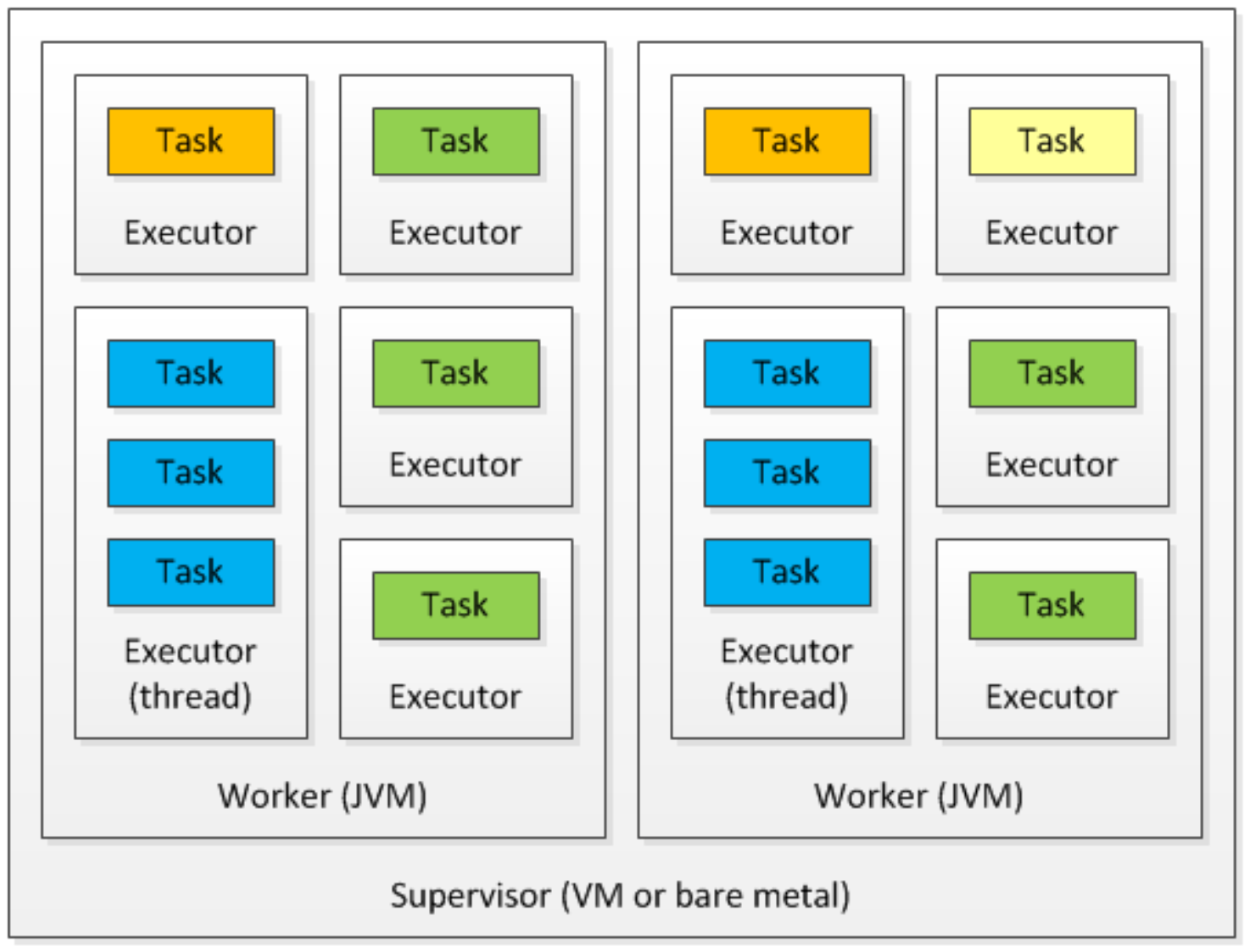}}}
    \caption{Master-Workers structure of the Spark runtime~(a) and Worker 
        hierarchy example in Storm~(b).}
    \label{fig:common-mw}
\end{figure}

\paragraph{The Master}
has total control over 
program execution, job scheduling, communications, failure management, 
resource allocations, etc. 
The master is the one that knows the semantic dataflow representing 
the current application, while workers are completely agnostic about the whole 
dataflow: they only obtain tasks to execute, that represent actors of the 
execution dataflow the master is running. 
It is only when the execution is effectively 
launched that the semantic dataflow is built and eventually optimized to obtain the best execution plan (\flink). With this postponed evaluation, the master creates what we called the parallel execution dataflow to be executed.
In \storm and \flink, the data distribution is 
managed in a decentralized manner, that is, it is delegated to each executor, 
since they use pipelined data transfers and forward tokens as soon as they are 
produced.
In \spark streaming, the master is the one 
responsible for data distribution: it discretizes the stream into micro-batches that 
are buffered into workers' memory.
The master generally keeps track of distributed 
tasks, decides when to schedule the next tasks, reacts to finished vs.\ failed 
tasks, keeps track of the semantic dataflow progress, and orchestrates  
collective communications and data exchange among workers. This last aspect is 
crucial when executing the so-called \emph{shuffle operation}, which implies a data exchange among 
executors. Whereas workers do not have any information about others, to 
exchange data they have to request information to the master and, 
moreover, specify they are ready to send/receive data. 

\paragraph{Workers}
are nodes executing the actor logic, namely, a 
worker node is a process in the cluster. Within a worker, a certain number of 
parallel executors is instantiated, that execute tasks related to the given 
application. Workers have no information about 
the dataflow at any level 
since they are scheduled by the master. Despite this, the different frameworks use different 
nomenclatures:
{in \spark, \storm and \flink 
cluster nodes are  decomposed into \emph{Workers}, \emph{Executors} and \emph{Tasks}.}
A Worker is a 
process in a node of the cluster, i.e., a \spark worker instance. A node may host 
multiple Worker instances. An Executor is a thread that is spawned in a Worker 
process and it executes Tasks, which are the actual kernel of an actor 
of the dataflow. Fig.~\ref{fig:common-mw}(b) illustrates
this structure in \storm, an example that would also be valid for \spark and \flink.

\subsection{Process-based Execution}
In \gtf, actors are 
effectively mapped to threads and possibly distributed on different nodes.
The cardinality of the semantic dataflow is preserved,
as each actor node is instantiated into one node, and the allocation is 
decided using a placement algorithm based on cost model optimization. 
The dataflow is distributed on cluster nodes and each node/Worker 
may host one or more dataflow \mbox{actors/Tasks}, that
internally implement data  parallelism with a pool of threads/Executors working 
on Tensors.  
Communication among actors is done using the send/receive paradigm, allowing workers 
to manage their own data movement or to receive data without involving the master 
node, thus decentralizing the logic and the execution of the application.

\section{Limitations of the \DF Model}
\label{sec:limit}

Reasoning about programs using the \DF model is attractive since it makes the program semantics independent from the underlying execution model.
In particular, it abstracts away any form of parallelism due to its pure functional nature.
The most relevant consequence, as discussed in many theoretical works about Kahn Process Network and similar models---such as \DF---is the fact that all computations are \emph{deterministic}.

Conversely, many parallel runtime systems exploit nondeterministic behaviors to provide efficient implementations.
For example, consider the Master-Workers pattern discussed in Section~\ref{sec:dpn}.
A naive implementation of the Master node distributes tasks to $N$ Workers according to a round-robin policy---task $i$ goes to worker $i\,(\text{mod } N)$---which leads to a deterministic process. An alternative policy, generally referred as \emph{on-demand}, distributes tasks by considering the load level of each worker, for example, to implement a form of load balancing.
The resulting processes are clearly nondeterministic, since the mapping from tasks to workers depends on the relative service times.

Non-determinism can be encountered at all levels of our layered model in 
Fig.~\ref{fig:stackmodel}.
For example, actors in \storm's topologies consume tokens from incoming streams according to a from-any policy---process a token from any non-empty input channel---thus no assumption can be made about the order in which stream tokens are processed.
More generally, the semantics of stateful streaming programs depends on the order in which stream items are processed, which is not specified by the semantics of the semantic \df actors in Section~\ref{sec:semDF}.
As a consequence, this prevents from reasoning in purely \DF---i.e., functional---terms about programs in which actor nodes include arbitrary code in some imperative language (e.g., shared variables).


\section{From Skeletons to Big Data, a Historical Perspective}
\label{sec:skel}

The need to exploit parallel computing at a high enough level of abstraction
certainly predates the advent (or the hype) of Big Data processing. In the
parallel computing and software engineering communities, this need
has been advocated years before by way of algorithmic
skeletons~\cite{Cole:1991} and design patterns~\cite{gamma-book},
which share many of the principles underlying the
high-level frameworks considered in previous sections. 
Conceptually, the tools we discussed through the paper
exploit \emph{Data Parallelism, Stream Parallelism}, or both.

Data Parallel patterns express computations in which the same kernel
function is applied to all items of a data collection, which include for instance Map and Reduce. They can be viewed as higher-order functions and
can be placed at the very top of our layered model from
Fig.~\ref{fig:stackmodel}, since they expose a declarative data
processing model (Section~\ref{sec:declarative}).

The ability to efficiently support
lists (tensors) transformations under a weakly ordered
execution models has been proved by Gorlatch's seminal work \cite{Gorlatch:1996:SEP:646669.701094}, 
which definitely influenced the design of MapReduce. Also, Map, Reduce and other data parallel
skeletons have been introduced and developed in many experimental
parallel programming frameworks, many of them described in the
survey from Gonz{\'a}lez-V{\'e}lez and Leyton~\cite{Gonzalez-Velez:2010:SAS}.

In Big Data oriented tools, data parallelism is
founded on the MapReduce
paradigm~\cite{dean2008mapreduce}%
,
a specific composition of the data parallel \emph{Map} and \emph{Reduce}
operations with an implicit intermediate shuffling/grouping phase,
which is made explicit in the frameworks we examined---e.g., \OP{groupByKey} in Spark,
\OP{groupBy} in Flink, \OP{fieldsGrouping} in Storm.
Such shuffling/grouping had not been addressed by parallel skeletons or patterns.

Stream Parallel patterns express computations in which data streams flow through a network of processing units.
It is another key parallelism exploitation
pattern, from the first high-level approaches to parallel computing,
such as the P3L language~\cite{orlando-grosso}, to more recent frameworks,
such as FastFlow~\cite{ff:acc:europar:11}.
This model, enriched with Control-Parallel patterns such as \OP{If}
and \OP{While}, allows to express programs through arbitrary graphs,
where vertexes are processing units and edges are network links. 
In this setting, Stream Parallel patterns represent pre-built,
nestable graphs, therefore they expose a topological data processing
model (Section~\ref{sec:topological}). 

In the Big Data context,
stream parallelism is interpreted in a quite
primitive way considering only basic management of data streams, such
as micro-batching, which turns stream parallelism into data
parallelism, and processing of independent data streams. More
advanced usage of streams can also be found in~\cite{dem:ijpp:2016}.

Parallel patterns can be composed to express arbitrarily complex data processing programs.
We already informally showed they subsume both declarative and topological data-processing models.
Moreover, \gtf can be regarded as a particular case of the two-tier
composition paradigm advocated in~\cite{opencl:ff:ispa:15}, in
which Data Parallel patterns are embedded only into Stream Parallel
patterns, thus resulting into a hybrid declarative/topological model.

\section{Conclusion}
\label{sec:conc}
In this paper, we showed how the \DF model  can be used to describe \BD analytics tools,
from the lowest level---process execution model---to the highest one---semantic \DF. The \DF model is expressive enough to 
represent computations in terms of batch, micro-batch and stream processing. 
With this abstraction, we showed that \BD analytics tools have similar 
expressiveness at all levels and we proceeded with the description of a layered model
capturing different levels of Big Data applications,
from the program semantics to the  execution model.
We also provided an overview of some well-known
tools---\spark, \flink, \storm and \gtf---by analyzing their semantics and 
mapping them to the proposed \DF-based layered model. With this work, we aim at giving 
users a general model to understand the levels underlying all the analyzed tools. 
Finally, we also described how the skeleton-based model provides an alternative but similar abstraction,
briefly describing some of most common parallel patterns that easily describe data parallel 
patterns used for \BD analytics. As future work, we plan to implement 
a model of \BD analytics tools based on algorithmic skeletons, on top of the 
FastFlow library~\cite{ff:acc:europar:11}.

\begin{acknowledgements}
This work was partly supported by the EU-funded project TOREADOR
(contract no.\ H2020-688797),  the EU-funded project Rephrase (contract no.\ H2020-644235),
and the 2015--2016 IBM Ph.D.\ Scholarship program. 
\end{acknowledgements}

\bibliographystyle{spmpsci}

\end{document}